\documentclass[a4paper,amsmath,amssymb]{jpconf}
\usepackage{graphicx}
\usepackage{amsmath, amsthm, amssymb}
\begin{document}
\title{Relation of classical non-equilibrium dynamics and quantum annealing}

\author{Hidetoshi Nishimori}

\address{Department of Physics, Tokyo Institute of Technology,
Oh-okayama, Meguro-ku, Tokyo 152-8551, Japan
}

\begin{abstract}
Non-equilibrium dynamics of the Ising model is a classical stochastic process whereas
quantum mechanics has no stochastic elements in the classical sense.
Nevertheless, it has been known that there exists a close formal relationship
between these two processes. 
We reformulate this relationship and use it to compare the efficiency of
simulated annealing that uses classical stochastic processes and quantum
annealing to solve combinatorial optimization problems.
It is shown that classical dynamics can be efficiently simulated by
quantum-mechanical processes whereas the converse is not necessarily true.
This may imply that quantum annealing may be regarded as a more powerful tool
than simulated annealing for optimization problems.
\end{abstract}
\section{Introduction}
\label{sec:Introduction}

Quantum annealing (QA) is a metaheuristic for combinatorial optimization problems.
In the words of physics, the task is to find the ground state of the Ising model 
with generally complicated interactions \cite{KN,STR,DC,SIC}.
QA uses quantum fluctuations realized by
transverse fields to search for the solution in the phase space.
A restricted version of QA is adiabatic quantum computation, in which
the system is supposed to follow the instantaneous ground state 
of the time-dependent Schr\"odinger equation \cite{EF}.
QA may be regarded as the quantum-mechanical alternative to
simulated annealing (SA) \cite{KGV,AK}, in which
classical thermal fluctuations play the role of quantum fluctuations in QA.
It is an interesting problem how QA can be better (or worse) than SA.

Recent years have observed high activities in this area, most of which
indicate some kind of supremacy of QA over SA \cite{KN,STR,DC,Kadowaki}.
Experimental approaches have also become available using the D-Wave machine
\cite{DW1}-\cite{DW18}.
Experimental data suggest that quantum effects indeed play roles
at least to some extent.

Pioneering examples of numerical data, obtained from direct numerical integration of
the Schr\"odinger equation and quantum Monte Carlo simulations,
are found in the thesis of Kadowaki \cite{Kadowaki}.
It has been found that the probability of the system being in the true
ground state is larger in QA than in SA for spin glass cases and travelling
salesman problems.
It has also been shown that the achieved values of the final energy are
lower in QA than in SA.

From the analytical side, theorems have been proven that give a
sufficient condition for the time dependence of the strength of quantum fluctuations,
{\it i.e.} the annealing schedule, such that the system reaches the ground state
in the infinite-time limit \cite{MN,MN2,MNMP,MoritaThesis}.
The resulting condition turns out to give a faster reduction of the control parameter
in QA than in SA.
It should nevertheless be remembered that this result does not mean that
a given difficult problem can be solved efficiently by the quantum method.
Theoretical supremacy of QA is usually quantitative,
not qualitative, as far as the sufficient condition for the worst case is concerned.

One of the interesting questions is whether or not QA can be efficiently simulated
on a classical computer.
The above-mentioned sufficient condition indicates that the amplitude of
the transverse field should be reduced as a function of time
in the same manner in the natural
Schr\"odinger dynamics and in the master equation dynamics corresponding to
quantum Monte Carlo simulations \cite{MN,MN2,MNMP}.
This is a surprising result because the Schr\"odinger dynamics is
completely different from the classical stochastic dynamics
of Monte Carlo processes.
The result may be understood to suggest that QA can be efficiently
simulated on a classical computer.
We should, however, be very careful
because, first, quantum Monte Carlo operates at low but finite temperatures
whereas the Schr\"odinger dynamics is a zero-temperature process.
Second, quantum Monte Carlo is a stochastic sampling
of the classical master equation.
Directly solving the master equation consumes an exponential amount of
time and memory on a classical computer, which is the same situation
as in solving the Schr\"odinger equation directly on a classical computer.

In the present paper, we try to shed some light on this problem.
Following \cite{NTK}, we show that
an ingenious mapping between classical stochastic dynamics and a quantum 
system allows us to derive a conclusion that simulated annealing may be
efficiently simulated by quantum annealing but the converse is not
necessarily true.

\section{Classical to quantum mapping}

Let us first consider the classical dynamics of the Ising model described by
the Hamiltonian $H_0(\sigma)$, where $\sigma$ stands for the set of
$N$ Ising spins, $\sigma=\{\sigma_1, \sigma_2,\cdots ,\sigma_N\}$.
We have added a subscript 0 to the Hamiltonian to distinguish this classical
Ising Hamiltonian from the quantum Hamiltonian defined later.
The master equation for the probability $P_{\sigma}(t)$ that the system is
in the state $\sigma$
\begin{equation}
\frac{dP_{\sigma}(t)}{dt}=\sum_{\sigma}W_{\sigma\sigma'}P_{\sigma'}(t),
\label{masterequation}
\end{equation}
where $W_{\sigma\sigma'}$ denotes an element of the transition matrix,
defines the classical stochastic dynamics.
The transition probability is a function of the temperature, which is
implicit in the above notation.
The temperature is fixed for the moment.
The transition matrix has non-positive right eigenvalues
\begin{equation}
W\psi^{(R,n)}=-\lambda_n \psi^{(R,n)}
\end{equation}
with $\lambda_0=0$ and $0>-\lambda_1>-\lambda_2>\cdots$.
The leading eigenvector $\psi^{(R, 0)}$ corresponds to the equilibrium state,
and other eigenvectors represent relaxing modes.
The solution to the master equation (\ref{masterequation}) has a general expression
as
\begin{equation}
P_{\sigma}(t)=P_{\sigma}^{\rm eq}+a_{\sigma}^{(1)} e^{-\lambda_1 t}
+a_{\sigma}^{(2)} e^{-\lambda_2 t}+\cdots,
\end{equation}
where $P_{\sigma}^{\rm eq}$ is the equilibrium Gibbs-Boltzmann distribution.
This expression makes it clear that the relaxation time toward equilibrium
is the inverse of the leading non-vanishing eigenvalue,
\begin{equation}
\tau_{\rm relax}=\frac{1}{\lambda_1}.
\end{equation}

The relaxation time diverges at a transition point $T=T_{\rm c}$.
If the transition is of second order, the relaxation time at $T=T_{\rm c}$
diverges polynomially as the system size $N$ increases,
\begin{equation}
\tau_{\rm relax}\Big(=\frac{1}{\lambda_1}\Big)\propto N^a~(a>0),
\end{equation}
whereas the divergence is usually exponential at a first-order transition,
\begin{equation}
\tau_{\rm relax}\Big(=\frac{1}{\lambda_1}\Big)\propto e^{bN}~(b>0).
\end{equation}
This means that the energy gap between the leading eigenvalue $\lambda_0=0$
and the next eigenvalue $\lambda_1$ closes polynomially at a second-order
transition point and exponentially at a first-order transition.

If we define a $2^N\times 2^N$ matrix $H$ with element \cite{NTK,Henley,Castel}
\begin{equation}
H_{\sigma\sigma'}=-e^{\beta H_0(\sigma)/2} W_{\sigma\sigma'}e^{-\beta H_0
(\sigma')/2}, \label{WtoH}
\end{equation}
where $\beta$ is the inverse temperature,
then the detailed balance condition $W_{\sigma\sigma'}P_{\sigma'}^{{\rm eq}}
=W_{\sigma'\sigma}P_{\sigma}^{{\rm eq}}$
imposed on the transition matrix
guarantees that $H$ is a symmetric matrix,
\begin{equation}
H_{\sigma\sigma'}=H_{\sigma'\sigma}.
\end{equation}
Since all matrix elements of $H$ are real, this is a Hermitian matrix
and thus can be regarded as the Hamiltonian of a quantum system.

The Hamiltonian $H$ shares the eigenstates and eigenvalues with $W$,
up to trivial factors,
\begin{equation}
H\phi^{(n)}=\lambda_n \phi^{(n)},~\phi^{(n)}=e^{\beta H_0/2}\psi^{(R,n)}
\end{equation}
as can be verified from the definition (\ref{WtoH}).
Since the energy gap between the ground and first excited states
$\Delta =\lambda_1$ is common to both systems (the classical dynamical
system and the quantum system), these two systems share the existence and properties
of a phase transition.
In particular, the energy gap of the quantum system closes polynomially
as a function of the system size at a second-order quantum transition
and exponentially at a first-order quantum transition if the same
is true for the classical case.

It is to be noticed that this classical-to-quantum mapping
preserves the spatial dimension of the system.
It is also important to remember that the definition (\ref{WtoH})
leads to short-range interactions in $H$ if the same is true for $H_0$
as can be verified from the definition (\ref{WtoH}).
A simple example is the one-dimensional ferromagnetic Ising model
\begin{equation}
H_0(\sigma)=-\sum \sigma_j \sigma_{j+1},
\end{equation}
from which the following quantum Hamiltonian is derived under
the heat-bath dynamics \cite{NTK},
\begin{equation}
H=-\frac{1}{2}\sum_j \sigma_j^z \sigma_{j+1}^z -\frac{1}{2\cosh 2\beta}
\sum_j (\cosh^2 \beta -\sinh^2 \beta~\sigma_{j-1}^z\sigma_{j+1}^z)\sigma_j^x.
\end{equation}

\section{Quantum to classical mapping}

The converse mapping from a quantum Hamiltonian to classical dynamics
can be formulated as follows \cite{NTK,Henley,Castel}.
Consider a quantum spin system that can be expressed in the basis
of eigenstates of $\{\sigma_j^z\}_{j}$ as $H_{\sigma\sigma'}$.
It is necessary to impose the conditions of negative semi-definiteness
of off-diagonal elements
\begin{equation}
H_{\sigma\sigma'}\le 0~(\sigma\ne\sigma')
\label{Hoffdiagonal}
\end{equation}
and positive semi-definiteness of the eigenvalues,
\begin{equation}
H\phi^{(0)}=0,
\label{Hdefiniteness}
\end{equation}
where $\phi^{(0)}$ is the ground-state eigenvector.
The latter condition can always be satisfied by a shift of the
energy standard.
Under these conditions, the matrix $W$ defined by
\begin{equation}
W_{\sigma\sigma'}=-e^{-H_0(\sigma)/2}H_{\sigma\sigma'}e^{H_0(\sigma')/2}
\end{equation}
can be regarded as the transition matrix of a classical stochastic
process of the Ising model $H_0(\sigma)$ whose elements are defined by
\begin{equation}
H_0(\sigma)=-2\log \phi_{\sigma}^{(0)}.
\label{ptoH}
\end{equation}
This last definition is valid because the ground-state eigenvector
$\phi^{(0)}$ has all its elements positive
\begin{equation}
\phi_{\sigma}^{(0)}>0
\end{equation}
due to the Perron-Frobenius theorem that applies to matrices with
the properties described in equation (\ref{Hoffdiagonal}).
Notice that the prescription (\ref{ptoH}) has been inspired by the relation
\begin{equation}
\phi_{\sigma}^{(0)}=e^{-\beta H_0(\sigma)/2}
\end{equation}
in the classical-to-quantum mapping.
It is also interesting that the present quantum-to-classical mapping
does not change the spatial dimension in contrast to the Suzuki-Trotter
decomposition or the path-integral formulation of quantum mechanics,
by which the effective spatial dimension increases by one.

The classical Ising Hamiltonian $H_0(\sigma)$ defined by equation
(\ref{ptoH}) has in general many-body long-range interactions
with $2^N$ coefficients
\begin{equation}
H_0(\sigma)={\rm const}-\sum h_j \sigma_j -\sum J_{ij}\sigma_i \sigma_j
-\cdots -J_N \sigma_1 \sigma_2 \cdots \sigma_N.
\label{qtocH0}
\end{equation}
The reason is that equation (\ref{ptoH}) must be satisfied by
all possible values of $\sigma=\{\sigma_1,\sigma_2 ,\cdots , \sigma_N\}$,
and there are $2^N$ of them.
Hence $H_0(\sigma)$ should have $2^N$ coefficients in front of
various products of spin variables as they appear in equation (\ref{qtocH0}).
We set up $2^N$ linear equations for these coefficients,
the solution of which gives, in general, non-vanishing values for
all coefficients.
This is in sharp contrast to the classical-to-quantum mapping, in which
short-range interactions are mapped to short-range interactions.
Since the many-body long-range interactions in equation (\ref{qtocH0}) are hard
to simulate, we may conclude that classical simulation of the quantum system
is not efficient.

\section{Conclusion}

We have reformulated the mapping between classical stochastic dynamics of
the Ising model and a quantum system in stationary state.
It has been shown that classical stochastic dynamics with fixed temperature
can be efficiently simulated by a quantum system whereas quantum systems cannot
necessarily be efficiently simulated by classical dynamics as long as
the present quantum-to-classical mapping is concerned.
The present results indicate that simulated annealing, if the temperature
is changed slowly, may be efficiently simulated by quantum annealing if
a control parameter in the quantum system, typically the transverse field,
is changed slowly or adiabatically.
The converse process to simulate quantum annealing with a slow change of a
control parameter by simulated annealing with a slow decrease of temperature
is not necessarily efficient.
It may be premature to conclude only from this result that quantum annealing
is superior to simulated annealing because other formulations
of classical-quantum correspondence may be possible in which the present
conclusion does not hold.
It is nevertheless important to have established that the present
method yields asymmetric correspondence between classical stochastic dynamics
and quantum mechanics.

\ack
This was done in collaboration with Junichi Tsuda and Sergey Knysh.

\end{document}